\documentclass{pasj00}
\draft
\usepackage{color}

\begin{document}
\SetRunningHead{Author(s) in page-head}{Running Head}

\title{Study of infrared excess from circumstellar disks in binaries with {\it Spitzer}/IRAC}

\author{Yusuke \textsc{Itoh}, Misato \textsc{Fukagawa}, Hiroshi \textsc{Shibai}, Takahiro \textsc{Sumi}, and Kodai \textsc{Yamamoto}}
\affil{Department of Earth and Space Science, Graduate School of Science, Osaka University, 1-1 Machikaneyama, Toyonaka, Osaka 560-0043 Japan.}
\email{itoh@iral.ess.sci.osaka-u.ac.jp, 
misato@iral.ess.sci.osaka-u.ac.jp, 
shibai@iral.ess.sci.osaka-u.ac.jp, 
sumi@iral.ess.sci.osaka-u.ac.jp, 
yamamoto@iral.ess.sci.osaka-u.ac.jp}


\maketitle

\begin{abstract}
The presence of excess emission at 3.6--8.0~$\mu$m was investigated in a sample of 27 binary systems located in two nearby star-forming regions, Taurus and Ophiuchus, by using {\it Spitzer}/Infrared Array Camera (IRAC) archival data. Angular (Projected) separations for the binaries are greater than 2$\arcsec$($\sim$280~AU), which allowed us to perform spatially resolved photometry of individual primary and secondary sources.
The measured occurrence of infrared excess suggests that binarity plays a role in the evolution of circumstellar disks, even at such wide binary separations. 
Most of the binaries have excess emission from both the circumprimary and circumsecondary disks, or show photospheric levels for both components at all four wavelengths of IRAC. 
On the other hand, four systems 
($17^{+11}_{-8}$\%, designated by ``mixed'' systems) exhibit excess emission from a single binary component.
This ratio is significantly smaller than that predicted by the random pairing of single stars, suggesting that circumprimary and circumsecondary disks are synchronously dispersed.
In addition, the excess frequencies (EFs) of primary and secondary sources with a projected distance of $a_{\rm p}\simeq280$--450~AU are $100^{+0}_{-17}$\% and $91^{+8}_{-18}$\%, respectively, and significantly higher than that of single stars ($70\pm5$\%).
We made a simple model describing the EF distribution as a function of the disk outer radius, $R_{\rm out}$. 
Comparisons with observations using the Kolmogorov-Smirnov test show that the observational data are consistent with the model when the $\rm{EF}\simeq1$ region is found at  $R_{\rm out}\sim30$--100~AU.
This disk radius is smaller than that typically estimated for single stars. 
The high EF of circumstellar disks with these radii may indicate a prolonged lifetime of dust in binary systems possibly because smaller disks counteract mass loss by photoevaporation. 
\end{abstract}

\section{Introduction}
The ubiquity of binary stars is widely known.
More than 60\% of main-sequence solar-type stars (e.g., \cite{abt76,duq91}) and $\sim$30--40\% of M-type stars have been reported as multiple systems (e.g., \cite{fis92,del04}).
In particular, the significant progress of the multiplicity survey of solar-type stars was achieved by \citet{duq91} based on the radial velocity measurements. 
\citet{rag10} updated and extended their results using an increased sample with long-baseline interferometry and speckle interferometry. 

Numerous young binaries have also been detected in nearby star-forming regions (e.g., \cite{ghe93,sim93}).
Observational study of young binaries is particularly important for understanding how most stars acquire their planetary systems. 
There are notable observations on circumstellar disks in binaries, although their evolutionary process is less clear than in single stars because sufficient data have not been accumulated. 
Circumprimary and circumsecondary disks around SR~24 in Ophiuchus have been directly imaged in the near-infrared with the high spatial resolution, enabling a detailed comparison with theoretical predictions of disk formation and evolution in binary systems \citep{art94, may10}. 
On the other hand, excess emission in infrared and millimeter regions and line emission as a diagnostic of accretion have been statistically analyzed to investigate the evolution of circumstellar disks in binaries (e.g., \cite{har12}). 
Recent ALMA observations yielded several new detections of low-mass circumsecondary disks thanks to the high sensitivity, and the increasing sample size helps to determine the disk population for each binary component \citep{ake14}.

In infrared and optical, consistent results have emerged from the statistical studies such as on the difference in disk frequency between primary and secondary stars, and the dependence of the disk lifetime on binary separation. 
\citet{whi01} investigated the nature of circumprimary and circumsecondary disks from their H$\alpha$ and $K-L$ colors. 
Because the lifetimes and accretion rates of circumprimary disks exceeded those of the secondary disks, they concluded that, in general, most of the material in a circumprimary disk is  preferentially supplied by the circumbinary disk.
\citet{mcc06} also studied the $K-L$, $K-N$ colors, and H$\alpha$ for binary systems. 
Their study showed that $\sim$10\% of the stars with the gas accretion and the excess in $K-N$ color were photospheric in $K-L$ color, indicating that their disks are in the middle of the inside-out evolution. 
These disks are harbored by the secondary stars with primary stars surrounded by disks, while only one of these is harbored by a primary star with a diskless secondary star; i.e., circumsecondary disks tend to disperse faster than circumprimary disks.
Similarly, \citet{mon07} used H$\alpha$ emission, $K-L$, and $K-N$ to investigate circumstellar disks in binaries. 
They found that there were fewer systems showing the gas accretion only onto the secondary stars than systems only accreting onto the primary stars. 
Therefore, they deduced that the lifespan of circumprimary disks exceeds that of circumsecondary disks.

Past studies have also revealed how the disk existence depends on binary separation (\cite{pra97,duc99b,whi01,har03,dae12}). 
\citet{cie09} showed that the lifetime of circumstellar disks in binary systems was significantly shorter at separations of 40~AU and below compared to disks of single stars. 
They investigated infrared excess from binaries in four nearby star-forming regions ($\sim$1--3~Myr) using the $[3.6]-[8.0]$ color obtained with {\it Spitzer}/IRAC without resolving the individual stellar components.
Accelerated dispersal of circumstellar disks in close binary systems is also suspected among the members of $\eta$~Chamaeleon, the ages of which are  $\simeq$8~Myr \citep{bou06}. 
Similar results have been recently reported by \citet{kra12}, who studied excess emission in Taurus binaries over a wide range of wavelengths, from near-infrared to millimeter. 
From these data, they calculated the disk frequency, also without resolving individual stellar components.
While $\sim$67\% of the close binaries ($\lesssim$40~AU) disperse their disks within $\sim$1~Myr, $\sim$80--90\% of single stars and wide binary systems ($\gtrsim$400~AU) retain their disks during $\sim$1--2~Myr.

Some properties of the circumstellar disks, such as the disk frequency (e.g., \cite{bou06,cie09,kra12}), gas accretion, and dust emission (e.g., \cite{whi01,dae12}), have been reported to be indistinguishable between wide binaries and single stars. 
While many studies indicate rapid disk dispersal in close binary systems, the dependence of the disk dissipation time scale on wide binary separation ($\gtrsim$100~AU) has been little investigated.
Especially, sample size was insufficient, and binary components were not spatially resolved at mid-infrared wavelengths where the presence of disks can be more reliably estimated than in near-infrared. 

In this paper, we discuss infrared emission from {\it each} stellar component of binary systems.
The analysis is based on spatially resolved photometry at four wavelengths, 3.6, 4.5, 5.8, and 8.0~$\mu$m, obtained from {\it Spitzer}/IRAC images.
The IRAC wavelengths correspond to a dust temperature of $\sim$360--800~K and a radial region within a few AU for M-type stars.
Thus, these wavelengths can probe the inner planet-forming radii in dusty disks.
In addition, {\it Spitzer}/IRAC bands located between $L$ and $N$ that are difficult to access with ground-based telescopes; therefore, new information such as the onset of infrared excess emission can be obtained.
Moreover, the unprecedented sensitivity of {\it Spitzer} enables the detection of lower-mass secondaries.
We collected binary systems with the projected separation $a_{p}$ of $2\farcs0$--$17\farcs0$, corresponding to $\sim$280--2400~AU at 140~pc.
Because the images of the primary and secondary stars overlap at projected separations near 2$\arcsec$, we analyzed each star by stellar profile fitting for the photometry.

The rest of the paper is organized as follows.
Section~\ref{sec:sample} describes our sample and the {\it Spitzer}/IRAC data.
Section~\ref{sec:photometry} explains the profile fitting photometry .
Section~\ref{sec:results} presents the measured frequency of excess emission in binary systems and the difference between binary systems and single stars.
The results are discussed in Sections~\ref{sec:binarity} and  summarized in  \ref{sec:discussion}.

\section{Sample and archival data}\label{sec:sample}
The full-width at half-maximum (FWHM) of a point source is 2$\arcsec$ at 8.0~$\mu$m for {\it Spitzer}/IRAC \citep{faz04}.
Given the array pixel scale of $1\farcs2$, we selected binary systems with projected separations greater than $2\farcs0$.
Thirty-three binaries of separation of $1\farcs95$--$17\farcs0$ in the star-forming regions of Taurus (distance 140~pc; \cite{tor07}, age 1--2~Myr; \cite{kra09a}) and Ophiuchus (distance 140~pc; \cite{loi08}, age 1~Myr; \cite{all06}; \cite{luh99}; \cite{pra03}) were extracted from the literature \citep{kra11,kra09b,duc07,rat05,hai04,hai02,whi01,koh98,rei93,lei93} (see also Table~1 for references on each object).

The companionship and membership in Taurus or Ophiuchus in our sample were examined using the literature because the sample may include suspicious secondary stars not physically associated to the primary, or older systems than 1--2 ~Myr.  
In Taurus, coevality ($d\Delta \log(t)$) between primary and secondary stars has been indicated to be $<$0.4~dex which is significantly smaller than 0.58~dex estimated from the random paring of single stars \citep{kra09a}. 
In our initial 33 binaries, the systems whose ages are known and coevality is larger than 0.4~dex are 2MASS~J04554757+3028077, HP~Tau/G2, and DK~Tau. 
The age difference of $d\Delta \log(t)\sim0.8$ was found for DK ~Tau between the primary and the secondary stars, but  
a possible systematic error
was implied in this age estimation  \citep{kra09a}. 
In addition, in order to explain the estimated size for the circumprimary disk, the presence of a companion is favored just by the distance of the secondary candidate. 
Therefore, we retained DK~Tau in our binary sample while excluded 2MASS~J04554757+3028077 and HP~Tau/G2.
In addition, RXJ~0437.4+1851, NTTS~040142+2150, and V1117~Tau systems are speculated to be older compared to the typical ages of Taurus based on the Li test, or to be unassociated to the Taurus region judging from the proper motions (\cite{ses08,wic00}).
Since contamination of these sources can affect the discussion of disk frequency at the age of 1--2~Myr, we eliminated these sources from our  sample.
For Ophiuchus, VSSG 23 was removed from our sample due to their non-coevality ($d\Delta \log(t)\sim1$; \cite{pra03}). 
Unfortunately, sufficient data on coevality and proper motions were not available for many of the Ophiuchus sources, and these systems remain to be included in the sample as binary candidates. 
As a result, six systems were eliminated from the initial 33 binaries, and 27 systems were analyzed in this paper. 

It should be noted that our sample includes triple and quadruple systems consisting of the close binaries with separations of $\sim$$0\farcs01$--$0\farcs1$ and the other widely separated components:  2E~1628.2-2423 \citep{mat89}, V1001~Tau \citep{whi01}, RXJ~0435.9+2352 \citep{koh98}, L1689~SNO2 \citep{rat05}, and 2MASS~J04251767+2617504 \citep{duc99}.
Because the close components should exhibit rapid disk clearing (\cite{mon07,cie09,kra12}), whether this parameter affects our discussion or not is explored in later sections.
Binarity in these systems, especially those that are widely separated, has yet to be confirmed by multi-epoch astrometry; therefore, many of these systems are candidates of binaries.
Table~1 summarizes the projected separations and spectral-types taken from literature for our sampled systems.

Since the closely spaced systems cannot be analyzed by aperture photometry, they were divided into individual components by profile fitting to the {\it Spitzer}/IRAC archival images, as described in the following section.
Fortunately, many of the binaries are located in regions surveyed in the ``From Molecular Cores to Planet-Forming Disks'' (c2d) legacy project \citep{eva03}.
The IRAC data were acquired from the NASA/IPAC Infrared Science Archives\footnote{http://irsa.ipac.caltech.edu/data/SPITZER/docs/spitzerdataarchives}. 
Following basic calibration, cosmic rays were removed by mosaicking the post-BCD data (IRAC Instrument Handbook\footnote{http://irsa.ipac.caltech.edu/data/SPITZER/docs/irac/iracinstrumenthandbook/}), and photometry was performed on the mosaicked post-BCD images.
The pixel scale of the post-BCD images was $0\farcs6$, two times as small as the native pixel size.
The data were obtained over two effective integration times: \verb|long| (10.4~sec) and \verb|short| (0.4~sec).
Photometry was generally carried out using the \verb|short|-exposure frames, because these frames yield better  fits to the stellar profile. 
The exceptions were the faint ($\lesssim$9 mag) binaries for which the \verb|long|-exposure images were used.

\section{PSF-fitting photometry}\label{sec:photometry}
The images of the primary and secondary stars overlapped at binary projected separations of $\sim$2$\arcsec$--4$\arcsec$.
Therefore, we resolved the primary and secondary stars by $\chi^{2}$ fitting using a stellar profile model, which determined six variables: the positions of the primary and secondary star ($x$ and $y$ for each star) and the scaling factors of the profile models. 
The six simultaneous equations were solved by the Newton--Raphson method.
The initial position of the primary star was set at the location of the brightest pixel and was adjusted to minimize $\chi^{2}$.
The location of the secondary star was assumed to match its relative position to the primary star, reported in the literature.
The location of the secondary star was then adjusted to minimize $\chi^{2}$.
The initial scaling factor for the profile model was determined to match the pixel intensity to its observed intensity at the initial stellar position.
To create the profile model, the images of five single stars near the binary system were averaged using the \verb|PSF| task in IRAF\footnote{IRAF is distributed by the National Optical Astronomy Observatories, which are operated by the Association of Universities for Research in Astronomy, Inc., under cooperative agreement with the National Science Foundation.}.
The uncertainty on the observed PSF used for the $\chi^{2}$ fitting was estimated as follows. 
First, each of the five single stars was fitted with the same fitting tool and profile model, 
using the standard deviation $\sigma$ stored in the 2D uncertainty image in the post-BCD dataset (\verb|*_unc.fits|) as a weight. Here, $\sigma$ contained dark noise, flat-fielding error, read-out noise, and Poisson noise. 
At each pixel, the difference between the observed value and the fitted model profile reached $\sim$10--20\%, exceeding those predicted from the uncertainty image by a few orders of magnitude. 
Thus, to further consider the possible uncertainty associated with the observed PSF, the mean of the residuals between the observed profile and the model was calculated from each of the five stars, then applied to the $\chi^2$ fitting as a new $\sigma$. 
Averaged stellar profile model, together with its $\sigma$, yielded plausible results ($\chi^{2}/{\rm d.o.f.}\sim1$).
We also confirmed the quality of the fitting in the residual image and that the location of each star was consistent in the different wavebands, even for closely spaced systems ($\sim$$2\arcsec$--$3\arcsec$).
The $\chi^{2}$ of all targets are summarized in Table~1.
The widely separated systems ($\gtrsim$4$\arcsec$) were fitted with the same tool to prevent any systematic error introduced by the fitting tool.

The averaged stellar profile could not be obtained for one system (TYC~1289-513-1), because no nearby single stars were found in the same field. 
This system was fitted by the template point response function (PRF) provided by the IRAC team.
Since the FWHM of the template function was always smaller than that of the observed PSF, it was convoluted with a Gaussian function. 
The values store in the post-BCD uncertainty image was used as $\sigma$ for the $\chi^{2}$ fitting. 
The resulting $\chi^{2}/{\rm d.o.f.}(\gtrsim1000)$ were unacceptably large, which should be plausible because the uncertainty associated with the observed PSF was ignored.
We verified the effect of variation in the stellar profile on the photometric results using systems that permitted an averaged stellar profile. The magnitudes measured between the Gaussian-convolved template and averaged stellar profile fitted to the data differed by $\lesssim$0.1~mag in all the IRAC wavebands.
Therefore, we concluded that the $\sigma$ in the uncertainty files were excessively small and responsible for the huge $\chi^{2}$, but the Gaussian-convolved template was reasonably fitted to the data.
          
The photometric error can be estimated using the revised $\sigma$ obtained from the empirical PSF constructed from the stellar profiles of five nearby single stars. 
The best-fitting scaling factor of the profile was varied until the reduced-$\chi^{2}$ changed by 1.17 ($\rm{d.o.f.}=6$), and the flux difference created by this process was considered to be an error in the measured magnitude.
The uncertainty estimates of several binaries were consistent within 10\% in all four wavebands.
Comparing the measured magnitudes with those obtained by Gaussian-convolved template fitting, we estimated a conservative photometric error of 0.14~mag in all the IRAC bands.

Photometry in some IRAC bands was precluded in three systems, because the primary and secondary stars were obscured by a large brightness contrast (at 3.6~$\mu$m for DoAr~24E and at 8.0 and 5.8~$\mu$m for HN~Tau~B) and by the overlap between the secondary position and the latent image caused by an exceptionally bright primary star (at 8.0~$\mu$m for 2E~1628.2-2423).
The measured magnitudes of all targets are summarized in Table~1. 

\begin{longtable}[hbp]{llllllll}
\caption{Characteristics of Sampled Binary Systems.}
\hline
\hline 
Systems	& Separation &	SpT. & [3.6] & [4.5] & [5.8] & [8.0] & $\chi^{2}/{\rm d.o.f.}$ (ch1--4)\\ 
		& ('')		     &         & (mag) & (mag) & (mag) & (mag) & \\  	
\hline
\endhead
\endfoot
\hline
\hline
\multicolumn{8}{l}{\hbox to 0pt{\parbox{170mm}{\footnotesize
NOTE.---For individual binary systems, spectral type (Column 3) and measured IRAC magnitudes (Columns 4--7) of the primary and secondary members are shown in the first and second row, respectively. 
The second column lists the projected separations of the systems. 
The eighth column lists the $\chi^{2}/{\rm d.o.f.}$ given by the profile fitting at each IRAC wavelength.
References are given in parentheses in Columns 2--3 (see below). 
The typical photometric error in the PSF fitting photometry is 0.14~mag. \\
\footnotemark[$^\dag$] Denotes a candidate of a triple or quadruple system. \\
\footnotemark[$^*$] Denotes a mixed system. \\
(1) \cite{whi04}; (2) \cite{whi01}; (3) \cite{kra11}; (4) \cite{cor06}; (5) \cite{pra03}; (6) \cite{kra09b};(7) \cite{wil05}; (8) \cite{kra09a}; (9) \cite{moni98}; (10) \cite{con10}; (11) \cite{kra07}; (12) \cite{duc07}; (13) \cite{koh98}; (14) \cite{rei93}; (15) \cite{rat05}; (16) \cite{hai04}; (17) \cite{hai02}; (18) \cite{gre95}; (19) \cite{luh99}; (20) \cite{wal94}; (21) \cite{ken95}; (22) \cite{duc99b}.}}}
\endlastfoot
	& 		     &     Taurus    &           &            &           &            &\\  \hline
JH~223                                                  & 2.07 (3)  & M2 (11)     & 9.10   & 8.88  & 8.39   & 7.88 & 1.0, 0.4, 0.7, 1.0 \\
                                                              &                  & M6.5 (11)  & 11.26 & 11.06 & 10.63 & 9.85 & \\
CoKu~Tau~3                                         & 2.1 (14)    & M1 (11)     & 7.51  & 6.98 & 6.51 & 5.67 & 0.8, 1.0, 1.5, 1.6 \\
                                                              &                  & M5 (11)     & 9.17  & 8.99 & 8.50 & 7.99 &  \\
DK~Tau                                                 & 2.3 (2)       & K9 (9)     & 6.21 & 5.77 & 5.50  & 4.87 & 1.0, 1.2, 1.2, 1.3 \\
                                                              &                  & M1 (9)     & 7.84 & 7.47 & 7.26  & 6.72 & \\
CIDA~9                                                 & 2.34 (3)   & M0 (11)     & 9.33 & 8.74 & 8.09 & 7.08 & 1.6, 1.0, 0.9, 1.3 \\
                                                              &                  & M2.5 (11)  & 11.49 & 11.01 & 10.70 & 9.82 & \\
HK~Tau                                                 & 2.34 (3)   & M1 (1)       & 7.81 & 7.49 & 7.14 & 6.62 & 0.7, 0.8, 0.4, 0.8 \\
                                                              &                  & M1 (1)       & 11.47 & 10.91 & 10.64 & 9.69 & \\                                       
IT~Tau                                                   & 2.39 (2)    & K3 (22)     & 7.54  & 7.16 & 6.82 & 6.32 & 1.8, 1.9, 0.8, 0.7 \\
                                                              &                  & M4 (22)     & 9.05  & 8.68 & 8.11 & 7.51 & \\
V1001~Tau$^\dag$                               & 2.62 (2)    & K8 (9)     & 6.69 & 6.27 & 5.88 & 5.33 & 0.8, 1.8, 1.1, 0.9 \\
                                                              &                  & M0 (9)    & 7.90 & 7.38 & 7.11 & 6.06 & \\
HN~Tau                                                 & 3.10 (6)  & K5 (4)       & 7.00 & 6.34 & 5.65 & 4.76 & 1.3, 3.0,---,--- \\
                                                              &                  & M4.5 (22)  & 11.11 & 10.51 & --- & --- & \\                                
V710~Tau$^*$                                       & 3.17 (2)    & M1 (21)     & 7.99 & 7.80 & 7.19 & 6.57 & 1.6, 4.1, 1.4, 3.1 \\ 
                                                              &                 & M3 (21)     & 8.39 & 8.45 & 8.26 & 8.29 & \\
2MASS~J04251767+2617504$^\dag$ & 3.4 (2)      & K7 (13)      & 8.44 & 8.38 & 8.34 & 8.31 & 2.9, 1.2, 1.8, 1.3 \\
                                                              &                 & M3 (13)     & 9.10  & 9.05 & 8.98 & 9.00 & \\
BBM92~14                                            & 4.06 (12)  & M2 (10)    & 6.55  & 5.84 & 5.32 & 4.57 & 1.2, 1.2, 1.6, 2.2 \\
                                                              &                 & M2 (10)    & 7.13  & 6.53 & 5.97 & 5.14 & \\
TYC~1289-513-1                                  & 6.87 (13)  & ---    & 8.68 & 8.67 & 8.57 & 8.47 & ---\\
                                                              &                 & ---    & 10.69 & 10.69 & 10.58 & 10.47 & \\
RXJ0435.9+2352$^\dag$                     & 11.3 (13)   & ---    & 8.79 & 8.78 & 8.76 & 8.63 & 9.2, 5.3, 2.3, 0.8 \\
                                                              &                  & ---    & 10.82 & 10.98 & 10.84 & 10.69 & \\                                           
GI~Tau                                                  & 13.14 (6) & K6 (21)     & 6.43 & 6.14 & 5.90 & 4.90 & 0.8, 1.0, 1.8, 1.0 \\
                                                              &                  & M0 (21)     & 6.83 & 6.38 & 5.81 & 4.82 & \\
FZ~Tau                                                  & 17.17 (6) & M0 (4)       & 6.28 & 5.75 & 5.28 & 4.62 & 1.5, 0.7, 1.6, 1.7 \\
                                                              &                  & K7 (21)      & 7.19 & 6.86 & 6.53 & 6.07 & \\ \hline
     & 		     &     Ophiuchus    &           &            &           &            &\\  \hline
DoAr~24~E                                             & 2.07 (15)  & G6 (7)       & --- & 5.03 & 4.62 & 3.88 & ---,1.8, 2.1, 1.4 \\
                                                              &                 & K4.5 (7)     & --- & 6.25 & 5.77 & 4.93 & \\
L1689~SNO2$^\dag$                           & 3.01 (15)  & M2 (10)      & 6.34 & 5.50 & 4.76 & 3.96 & --- \\
                                                              &                 & M2 (10)      & 8.04 & 7.72 & 7.37 & 6.76 & \\
WSB71a                                                & 3.56 (15)  & K2 (5)        & 6.62 & 6.08 & 5.46 & 4.75 & 1.7, 0.7, 2.4, 1.3 \\
                                                              &                 & M6 (5)        & 9.49 & 9.15 & 8.53 & 7.80 & \\
ROXRF~36                                           & 3.59 (15)  & ---     & 9.26 & 9.16 & 8.13 & 9.01 & 1.2, 0.8,---, 0.2\\
                                                              &                 & ---     & 10.09 & 9.96 & 9.80 & 9.77 & \\
WL~18                                                   & 3.62 (15)  & ---     & 8.67 & 8.00 & 7.84 & 7.35 & 2.0, 2.7, 1.5, 0.3 \\
                                                              &                 & ---     & 11.06 & 10.68 & 10.34 & 10.22 & \\
2MASS~J16264848-2428389                & 4.15 (16)  & ---     & 9.08 & 8.21 & 7.56 & 6.74 & 1.9, 3.3, 1.4, 1.2 \\
                                                              &                 & ---     & 10.38 & 9.71 & 8.94 & 8.25 & \\
2E~1628.2-2423$^\dag$$^*$               & 4.3 (14)    & G0 (20)   & 6.37 & 5.96 & 5.30 & --- & 2.0, 1.0, 2.7,--- \\
                                                              &                 & K5 (20)    & 7.68 & 7.59 & 7.50 & --- & \\
2MASS~J16262097-2408518$^*$        & 5.21 (15)  & M3 (4)     & 9.14 & 9.13 & 9.06 & 8.98 & 2.2, 1.0, 0.4, 0.2 \\
                                                              &                 & M7 (4)      & 10.89 & 10.53 & 10.10 & 9.37 & \\
2MASS~J16271757-2428562               & 8.55 (16)  & F7 (18)    & 7.82  & 6.95 & 6.47 & 6.49 & 5.2, 7.8, 6.0, 2.0 \\
                                                              &                  & M4 (19)   & 8.36  & 7.35 & 6.51 & 5.95 & \\
2MASS~J16263682-2415518               & 9.08 (12)   & M0 (3) &  8.17 & 7.85 & 6.29 & 5.27 & 4.1, 2.3, 2.7, 1.5 \\
                                                              &                  & M5 (7) &  10.25 & 10.23 & 9.74 & 9.14 & \\
UCAC2~21797671$^*$                         & 10.3 (14)   & M2 (7) &  8.83 & 8.48 & 8.23 & 7.72 & 0.8, 1.9, 1.7, 0.8 \\
                                                              &                  & M4.5 (7) &  10.07 & 10.04 & 9.96 & 9.87 & \\
2MASS~J16262404-2424480               & 10.47 (17) & --- &  6.43 & 4.83 & 4.18 & 3.44 & 1.6, 5.0, 2.9, 3.8 \\
                                                              &                   & ---  &  7.44 & 6.55 & 5.78 & 5.04 & \\    
\end{longtable}

\section{Results}\label{sec:results}
\subsection{Excess emission in IRAC bands}\label{sec:excess8}
\begin{figure}[htbp]
\begin{center}
\FigureFile(80mm,80mm){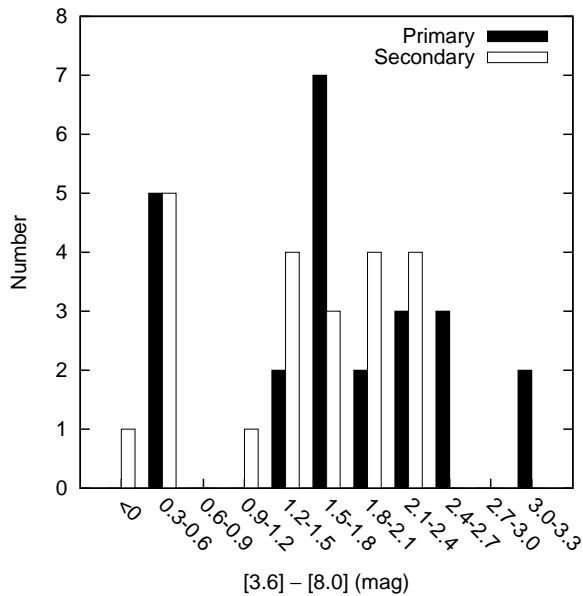}
\end{center}
\caption{Histogram of the color of $[3.6]-[8.0]$ shown by primary ({\it filled}) and secondary ({\it hollow}) sources. Class I sources (see Section 4.2) are not included in this plot. }\label{fig:hist}
\end{figure}

We first examined whether the emission of the individual components in the binary systems exceeds the photospheric level at 8.0~$\mu$m.
Figure~\ref{fig:hist} shows a histogram of the color $[3.6]-[8.0]$ distribution of the sources for which the photometry at 3.6 and 8.0~$\mu$m is available. 
The clear gap in the figure, occurring at $[3.6]-[8.0]\sim0.6$, is considered to divide the sources with and without excess at 8.0~$\mu$m.
The excess frequency (defined as (excess sources)/(excess sources $+$ non-excess sources), and hereafter denoted EF) of the primary and secondary stars were $79^{+5}_{-7}$\%(19/24) and $73^{+6}_{-7}$\%(16/22), respectively.

\begin{figure}
\begin{center}
\FigureFile(80mm,80mm){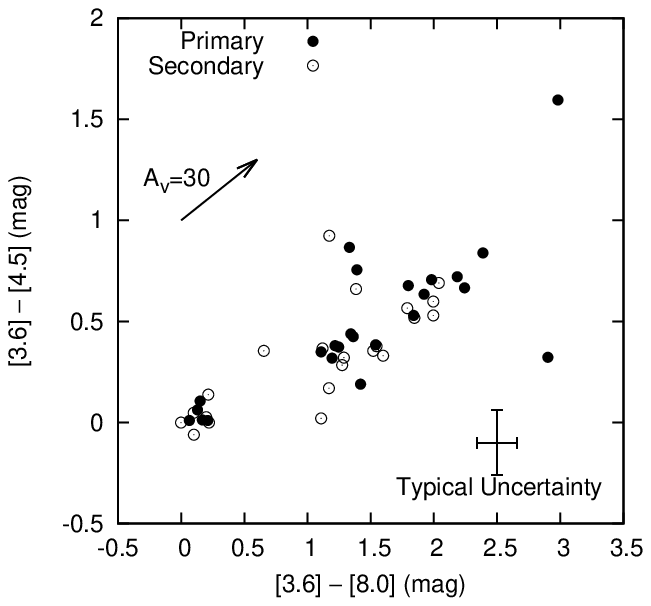}
\FigureFile(80mm,80mm){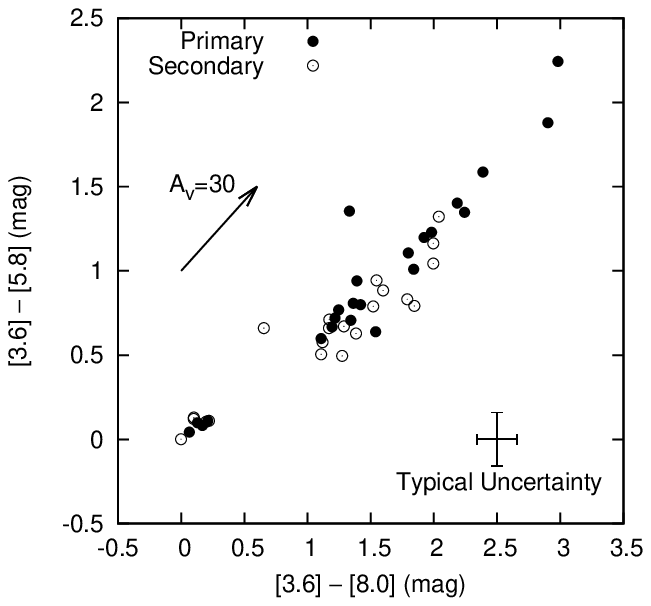}
\end{center}
\caption{
Color-color diagrams of $[3.6]-[4.5]$ ({\it left}) and $[3.6]-[5.8]$ ({\it right}) against $[3.6]-[8.0]$. 
{\it Filled circles} show primary stars and {\it hollow circles} indicate secondary stars. 
The extinction vector in the IRAC bands is based on the simple-fit formula reported by \citet{ind05}. Class I sources (see Section 4.2) are not included in this plot.
\label{fig:iraccolorcolor}}
\end{figure}

Figure~\ref{fig:iraccolorcolor} shows color-color diagrams of $[3.6]-[4.5]$ ({\it left}) and $[3.6]-[5.8]$ ({\it right}) against $[3.6]-[8.0]$. 
The objects with $[3.6]-[8.0] \lesssim 0.6$ are gathering near the origin of the diagrams, indicating that they are also photospheric at 4.5 and 5.8~$\mu$m. 
In contrast, the sources with $[3.6]-[4.5]\gtrsim0.2$ and $[3.6]-[5.8]\gtrsim0.4$, located in the correlation extending to the upper right part of the diagrams, tend to have excess emission at 8.0~$\mu$m ($[3.6]-[8.0]>0.6$) in addition to 4.5 and 5.8~$\mu$m. 
However, it should be noted that identification of the excess sources at 4.5~$\mu$m is difficult because no clear color break occurs. 
Although we can see a clear gap at $[3.6]-[5.8]$ color around 0.4, some excess sources overlap with the non-excess sources when considering the typical uncertainty of the $[3.6]-[5.8]$ colors.

The 8.0~$\mu$m excess was detected relative to 3.6~$\mu$m, rather than absolutely, based on the color $[3.6]-[8.0]$. In addition, the excess at 4.5 and 5.8~$\mu$m is sometimes difficult of identification by the color-color diagrams. Therefore, we additionally checked the presence of excess emission using the spectral energy distributions (SEDs). 
Fitting a stellar photosphere requires photometry in one or more optical bands in addition to $J$~band.
The photometric values in the optical bands ($B$, $V$, $R$, $I$) were taken from the USNO-B1 Catalog \citep{mon03}.
For the sources for which the resolved photometric values were unavailable from the catalog, we used the resolved data in optical bands from additional references \citep{zac05,tor06,gra07,her88}.
The $J$-band magnitudes were taken from the 2MASS All-Sky Catalog of Point Sources \citep{cut03}.
If these magnitudes were unavailable from the catalog, 
the resolved data in the $J$~band were extracted again from additional references \citep{kra07,con10,mon91,cha00}.
The spectral types were  available for many of our sample sources in previous studies (see Table~1).
Free parameters for the fitting are $A_{V}$ and the scaling factor of the photosphere, corresponding to the radius of the star and distance to the star-forming regions.
To model the stellar photosphere at high and low effective temperatures, we adopted the models of Krucz \citep{kur93} and AMES-Cond \citep{all01}, respectively.
For sources with known $A_{V}$ but unknown optical-band magnitudes, the quoted $A_{V}$ values were used to correct for the extinction in the SEDs \citep{fur11,cha00,wah10,eva09,cur11,mcc10,bon01}. 
The data required for the SED fitting were unavailable for 12 of the sources, and we do not discuss them below. 

The excess at 8.0~$\mu$m was estimated based on the SEDs for 37 sources with the available photometry at this wavelength. All the sources which appear at $[3.6]-[8.0]>0.6$ show the excess at 8.0~$\mu$m in their SEDs, while none of them at $[3.6]-[8.0]<0.6$ exhibit an excess emission. 
We also confirmed that 34 out of the 37 sources ($92^{+4}_{-6}$\%) showed or did not show excess at both 3.6 and 8.0~$\mu$m. 
Six sources (ROXRF~36~A and B, TYC~1289-513-1~A and B, RXJ0435.9+2352~A and B) were  excluded in the above calculation, because they lacked spectral-type data.
However, these sources were assumed to be non-excess sources in all the IRAC bands, since they were classified as Class~III based on $K-[8.0]$ data (see next subsection).
If the six sources are included, 40 out of the 43 selected sources ($93^{+3}_{-6}$\%) either show or do not show excess at both 3.6 and 8.0~$\mu$m. 

The occurrence of excess emission and the relation with the 8.0~$\mu$m excess were also examined at 4.5~$\mu$m and 5.8~$\mu$m in the SEDs. Thirty nine out of 54 sources were used since the photometric data were obtained for them at these wavelengths and 8.0~$\mu$m. Note that the sample for 5.8~$\mu$m is different from that of 4.5~$\mu$m, although the sample size is the same. 
At 4.5~$\mu$m, 37 out of the 39 sources ($95^{+3}_{-5}$\%) showed excess or non-excess at both wavelengths (4.5~$\mu$m and 8.0~$\mu$m), and at 5.8~$\mu$m, 38/39 sources ($98^{+2}_{-4}$\%) present the same characteristics. 
These results indicate that the presence or non-presence of excess emission at 8.0~$\mu$m behaves similarly at 3.6, 4.5, and 5.8~$\mu$m.
The exceptions are three sources that show excess at 8.0~$\mu$m but not at the shorter wavebands. 
2MASS~J16263682-2415518~B, V~710~Tau~A, and JH~223~A begin to exhibit significant excess from $\sim$8.0, 5.8, and 4.5~$\mu$m, respectively. 
These three sources are further discussed in the appendix.

\subsection{Our sample of disk-bearing stars}\label{sec:assignment} 
Following the results in the previous subsection, we evaluate the existence of a circumstellar disk through the color of $[3.6]-[8.0]$ since it shows the most distinct boundary between excess and non-excess stars.
However, the magnitudes at 8.0~$\mu$m were unavailable for two sources (2E~1628.2-2423 and HN~Tau~B); thus 
we attempted to speculate whether the sources lacking in the photometric data at 8.0~$\mu$m had excess at this wavelength (i.e., $[3.6]-[8.0] > 0.6$). 
2E~1628.2-2423~A shows significant excess at 3.6--5.8~$\mu$m, as confirmed from the SED, thus it is plausible to categorize it as an 8.0~$\mu$m excess source. 
On the contrary, 2E~1628.2-2423~B exhibited no excess at 3.6--5.8~$\mu$m. 
Considering the rarity of sources that first show excess at 
8~$\mu$m ($\sim$6\%), 
this object was presumed as a non-excess source at 8.0~$\mu$m. 
The value of $[3.6]-[4.5]=0.41$ for HN~Tau~B (M4.5; \cite{kra09a}) is significantly higher than the photospheric color of an M5-type star ($-0.08$; \cite{luh10}).
Therefore, this source most likely has 8.0~$\mu$m excess.
Note that the recent ALMA observations failed to detect the secondary source in dust  continuum at 850~$\mu$m and 1.3~mm \citep{ake14}, but the secondary  indicated the H$\alpha$ emission and was classified as a classical T Tauri star (e.g., \cite{duc99}).

Photometry at 3.6~$\mu$m was not obtained for both components in DoAr~24~E (see Section~\ref{sec:photometry}), and hence the $[3.6]-[8.0]$ colors were unavailable. 
However, they generated the significant excess at 4.5--8.0~$\mu$m, as confirmed by the SEDs.
Therefore, they can be categorized as excess sources. 
In summary, in the statistical discussion in later sections, we regard 2E~1628.2-2423~A,  HN~Tau~B, and DoAr~24~E as excess sources and 2E~1628.2-2423~B as a non-excess star. 

Our sample includes Class~I sources that might still be surrounded by envelopes.
To simplify the discussion and to calculate the excess frequency for Class~II and~III sources only, we excluded the Class~I sources as follows. 
Classes were identified from the $K-[8.0]$ color, adopting the criterion of \citet{lad87}.
That is, sources satisfying $K-[8.0]>3.9$ were regarded as Class~I, $1.14<K-[8.0]<3.9$ as Class~II, and $-0.26<K-[8.0]<1.14$ as Class~III.
The $JHK$-bands magnitudes were taken from the 2MASS Catalog or additional references (\cite{kra07}; \cite{con10}; \cite{mon91}; \cite{cha00}).
For the sources lacking in the SED fitting (Section~\ref{sec:excess8}), the $A_{V}$ values for the color correlation were taken from previous studies (\cite{fur11}; \cite{kra09a}; \cite{wah10}; \cite{cha00}; \cite{eva09}; \cite{mcc10}; \cite{cur11}; \cite{bon01}).
As a result, 35 out of the 54 sources with both $K-[8.0]$ and $A_{V}$ data available were divided into their respective classes, and two sources (2MASS~J16271757-2428562~B and CIDA~9~A) were identified as Class~I.
The remaining 19 sources lacking $K-[8.0]$ or $A_{V}$ data were classified based on the $J-H~versus~H-K$ color-color diagram.
In this classification, we employed the $JHK$ colors of unreddened main sequence of \citet{koo83} \citep{lad92} and the locus of T~Tauri stars derived by \citet{mey97}. 
In addition, we assumed the interstellar reddening raw estimated by \citet{koo83}, \citet{bes88}, and \citet{mar90}.
As a result, one source (2MASS~J16262404-2424480~B) was classified as Class~I. 
The 51 sources or 24 systems not classified as Class~I were used for the analyses described in Section 5.

\section{Effect of binarity for IRAC excess and differences from single stars}\label{sec:binarity}
In this section, we show whether primary and secondary stars exhibit the same behavior regarding the presence of IRAC excess. In addition, we discuss the dependence of EF on binary separation and the difference of the EF between the binary systems and single stars. 

\subsection{Mixed systems}\label{sec:mix}
\begin{figure}
\begin{center}
\FigureFile(80mm,80mm){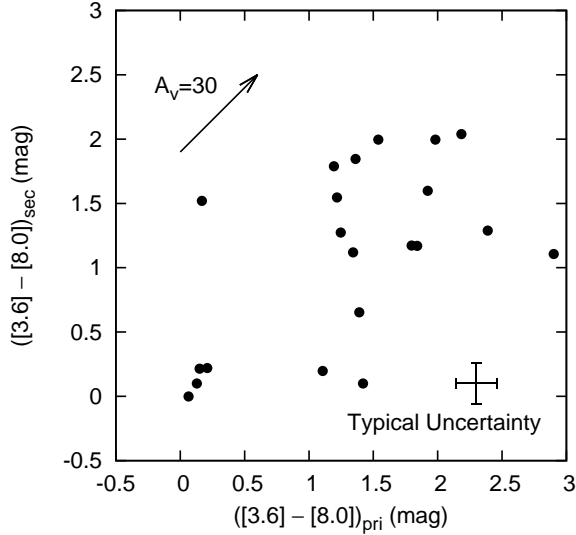}
\end{center}
\caption{$[3.6]-[8.0]$ color of secondary {\it versus} primary stars.  
The extinction vector in the IRAC bands is based on the simple-fit formula reported by \citet{ind05}. 
The typical error in the measured color is 0.20~mag. The sources without measurements of $[3.6]-[8.0]$ for both binary components are not included in the figure. 
\label{fig:pvss}}
\end{figure}

In Figure~\ref{fig:pvss}, the $[3.6]-[8.0]$ color of the secondary star is plotted against that of the primary star.
As described in the previous section, the clear color break at $\sim$0.6 divides the sources showing excess from those not showing excess.
Twenty systems out of the 24 binaries ($83^{+8}_{-11}\%$) show excess or non-excess associated with both components.
The remainder ($17^{+11}_{-8}$\%) show excess emission from only one of the binary components (``mixed'' binaries).
The number of systems showing excess solely from the primary and that solely from secondary source are three ($13^{+9}_{-6}$\%) and one ($4^{+7}_{-4}$\%), respectively.
To compare this mixed system ratio with the distribution of random sampling from a single star population, the EF of single stars was set to 70\% (see~Section~\ref{sec:single}).
The probability of obtaining the observed number of mixed systems is $C^{4}_{24}(0.42)^{4}(0.58)^{20}\simeq6\times10^{-3}$.
This significantly low probability implies that the disappearance of excess emission from circumprimary and circumsecondary disks is strongly correlated.
On the other hand, a binomial statistical analysis revealed no preference for which component is exhibiting excess; i.e., there is no significant difference between the ratio of systems showing excess solely from the primary source (3/4; $75^{+20}_{-36}\%$) and from the secondary source (1/4; $25^{+36}_{-20}\%$). 
The errors stated here represent the 1$\sigma$ confidence interval containing the central 68\% of the binomial distribution.

Our sample includes five triple or quadruple systems with close ($\lesssim0''.1\simeq15$~AU) binaries corresponding to the primary or secondary component.
Binaries with separations closer than $\sim$40 AU show a lower frequency of infrared excess compared with single or wide binary sources (e.g., \cite{mon07,cie09,kra12}). 
This suggests that these triples and quadruples should have a higher fraction of mixed pairs.
However, the EF does not significantly change when these five systems are excluded from the analysis. 
In this case, in 16 out of 19 systems ($84^{+8}_{-13}$\%), both members either show or do not show excess emission.
The number of mixed systems is 3 ($16^{+13}_{-8}$\% of the sample), and the conclusion is unaltered.

The four mixed systems are V710~Tau, 2MASS~J16262097-2408518, 2E~1628.2-2423, and UCAC2~217971. 
They can be in the evolutionary transition between disk-bearing and disk-less binaries. Based on the observed ratio of mixed systems and the ages of Taurus and Ophiuchus which are $\sim$$10^{6}$~yrs, the duration of the mixed phase can be speculated as $\lesssim$$10^{5}$~yrs. 
Alternatively, as mentioned above, it is known that the existence of a tight companion leads to a shorter timescale for disk clearing. Since very high angular-resolution observations have not been reported for the 4 binaries, we cannot exclude the possibility of these systems possessing undetected close ($\lesssim$10 AU) companions.

The rarity of mixed systems contradicts the concept of random pairing suggested in previous studies.
In their study of circumstellar disks in the Taurus region, \citet{whi01} reported that the disks at the separations wider than 210~AU are consistent with random pairing. 
This conclusion was inferred from the spatially resolved $K-L$ color and H$\alpha$ data.
More recently, \citet{dae12} suggested that random pairing occurs in the systems separated by more than 200~AU in the Orion Nebula Cluster. 
In contrast, our analysis indicates the correlation of primary and secondary components in the separation range of $\simeq$280--2400~AU.
Different sample size may account for the disagreement since our sample of 
24 systems is approximately three times greater than those of the previous works. 
Another potential source of disagreement is the disk indicators. 
In the above-mentioned studies, gas accretion and dust emission were probed from the innermost regions.
By contrast, the $[3.6]-[8.0]$ color traces dust emission from the farther, $\sim$1~AU region of the disk.

\subsection{Dependency of EF on separation}\label{sec:efvssep}
\begin{figure}
\begin{center}
\FigureFile(80mm,80mm){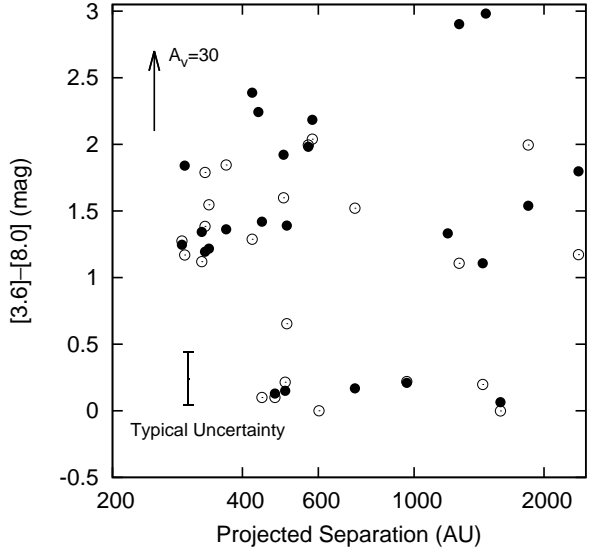}
\end{center}
\caption{$[3.6]-[8.0]$ {\it versus} binary separation for primary ({\it black circles}) and secondary ({\it unfilled circles}) stars.
Separation was converted to physical distance (in AU) using the distance to each star-forming region (140~pc). 
Note that the separation is the projected separation. 
The sources of excess emission are clustered around 280--450~AU. The sources without measurements of $[3.6]-[8.0]$ are not included in the figure.
\label{fig:i1i4vsap}}
\end{figure}

The $[3.6]-[8.0]$ color is plotted as a function of binary projected separation $(a_{\rm p})$ in Figure~\ref{fig:i1i4vsap}. 
The horizontal axis is converted from arc seconds to AU, assuming that each star-forming region is located at 140~pc.
In this figure, more closely separated binary companions ($a_{\rm p}\simeq280$--450 AU) tend to show excess emission. 
We verified the dependence of the EF on binary separation and  distinguished the regions exhibiting high and low EFs employing Fisher's exact test below.

\begin{figure}
\begin{center}
\FigureFile(78mm,80mm){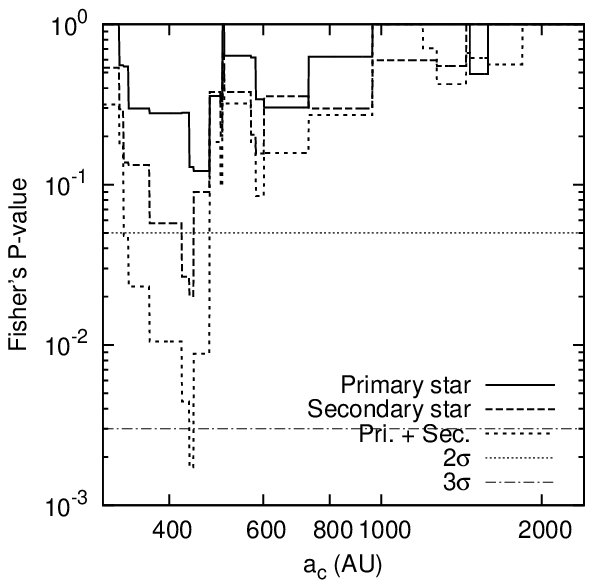}
\FigureFile(80mm,80mm){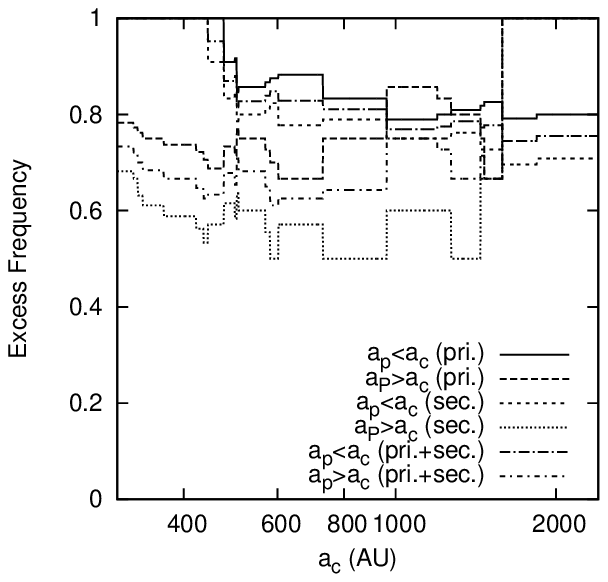}
\end{center}
\caption{
This figure shows the significance of EF differences between {\it closer} ($a_p < a_c$) and {\it wider} ($a_p > a_c$) binary systems.
{\it Left}: Fisher's $p$-value {\it vs.} $a_{\rm c}$ plane. The differences in EF were tested by Fisher's exact test between two groups of the binary samples, divided by whether the projected separation exceeds $a_{\rm c}$. The {\it solid}, {\it dashed}, and {\it dotted} line show the Fisher's $p$-value for primary, secondary, and primary+secondary sources, respectively.
The {\it horizontal-dotted} and {\it dashed-dotted} line indicate the 2$\sigma$ and 3$\sigma$ levels, respectively.
The minimum $p$-value occurs at $\sim$450~AU. 
At this radius, the significance of the EF difference between $a_{\rm p}<a_{\rm c}$ and $a_{\rm p}>a_{\rm c}$ is maximized ($2\sigma<p<3\sigma$ for each component and $>$3$\sigma$ for primary+secondary sources).
{\it Right}: The EF of each group bifurcates in all three sub-populations.
\label{fig:fishercvsw}}
\end{figure}

To identify the different EF regions, we split our sample into two groups based on whether the projected separation was larger or smaller than the separation $a_{\rm c}$ and assessed whether EF differed between the groups.
Figure~\ref{fig:fishercvsw} shows the results of Fisher's test applied to the three sub-populations; primary, secondary, and both components.
The {\it left} and {\it right} panels show Fisher's $p$-value and EF, respectively, as a functions of $a_{\rm c}$.
In the {\it left} panel, the $p$-value initially decreases at smaller $a_{\rm c}$, and is minimized at $a_{\rm c}\simeq450$~AU. 
This is because the sample size with $a_{\rm p}<a_{\rm c}$ increases while maintaining the high EF, whereas that of binaries with $a_{\rm p}>a_{\rm c}$ remains at $\sim$75\% as shown in the {\it right} panel.
Fisher's exact test clarified that the EFs of more closely and widely separated systems are different with a boundary at $\simeq$450~AU, at the significance of 1--2$\sigma$ for each binary component. 
Including both primary and secondary sources in the analysis, Fisher's test gave the significance of 2--3$\sigma$. 
It is notable that the EF is very high ($\sim$100\%) at $a_{\rm p}\lesssim450$~AU for each primary and secondary subpopulation, and naturally for the sample including both components ({\it right} panel of Figure~\ref{fig:fishercvsw}).

\subsection{Comparison of binary systems and single stars}\label{sec:single}
To compare the EFs between the binary systems with $a_{\rm p}\lesssim450$ AU and single stars, we first estimated the EF of single stars (${\rm EF}_{\rm single}$) in Taurus and Ophiuchus. 
In Taurus, the sample of 39 stars was adopted from \citet{kra12} because little contamination of tight binaries is expected given the unprecedented resolution ($\sim15$--20~mas; 2--3~AU) and sensitivity ($\Delta K\sim5$--6 mag at 40~mas;$\sim7$--15~$M_{J}$ at $\sim$6~AU) in their observations. 
The single stars in Ophiuchus were collected from \citet{cie09}. Unfortunately, the single nature of these stars is less clear compared to the Taurus stars and the sample can be contaminated by tight binaries. 
Since there is a possibility of EF lowered for single stars because of this contamination, we attempt to calibrate the EF for Ophiuchus stars based on the frequency of tight binaries found in Taurus, in the discussion later in this section.

The EF can be biased by stellar age. 
In fact, the disk frequency is well-correlated with the typical age of a star-forming region (\cite{mam09}).
Theoretically, the evolution of a pre-main sequence star at $\sim$1~Myr with a spectral type of G--M is  characterized by a decrease in luminosity at a nearly constant temperature (\cite{bar98,yi01}).
This means that the stellar luminosity can be substituted for the age; that is, the higher the luminosity, the younger the star and {\it vice versa}.
Thus, we checked whether the age dependence of ${\rm EF}_{\rm single}$ existed in our sample through the luminosities, which were estimated from the SEDs. 
Sufficient data for the SED fitting were available for 31 Taurus and 62 Ophiuchus sources.

To examine the dependence of EF on luminosity, the Taurus sample was divided into two subpopulations of spectral types of G0--K7 and M0--M4, which are comprised of 12 and 19 sources, respectively.
Each subpopulation was then divided by luminosity into two groups of approximately the same size.
The ${\rm EF}_{\rm single}$ was determined from the $[3.6]-[8.0]$ colors. 
The magnitudes at these wavelengths were obtained from IPAC/Gator\footnote{http://irsa.ipac.caltech.edu/applications/Gator/}. 
When the data were unavailable, the photometric results were extracted from the literature \citep{wah10,luh10}.
A systematic difference between the above-cited magnitudes and those measured with our PSF-fitting tool do not affect the EF comparison between single and binary systems, because in both cases, the $[3.6]-[8.0]$ distribution is clearly separated into two groups: with and without the color excess.
Using the same criteria ($[3.6]-[8.0]=0.6$), the EFs of four subpopulations  were indistinguishable (EF$\simeq$0.7--0.9), indicating that ${\rm EF}_{\rm single}$ does not significantly depend on luminosity. 
Thus, we infer that the ${\rm EF}_{\rm single}$ is independent of stellar age. 
When we added seven M0--M4 stars without sufficient observations for SED fitting, the 38 sources of Taurus gave an EF of 74\% (28/38). 
The same analysis was applied to the Ophiuchus sample, and a similar result was obtained about the EF dependence on the stellar luminosity. To equalize the comparison between the single and binary samples, we discarded one object of the spectral type later than M6 because our binary sample consists of $<$M6 members. The EF was calculated as 56\% (34/61) for the Ophiuchus cample. 
The final sample of single stars consists of 38 in Taurus and 61 sources in the Ophiuchus regions. 

\begin{figure}
\begin{center}
\FigureFile(78mm,80mm){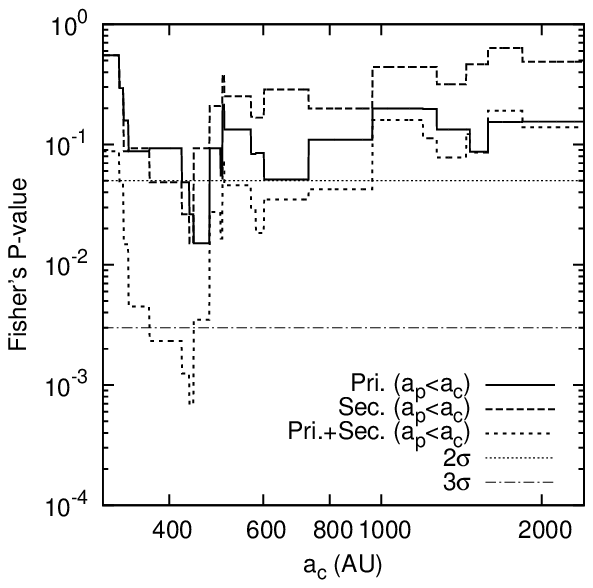}
\FigureFile(80mm,80mm){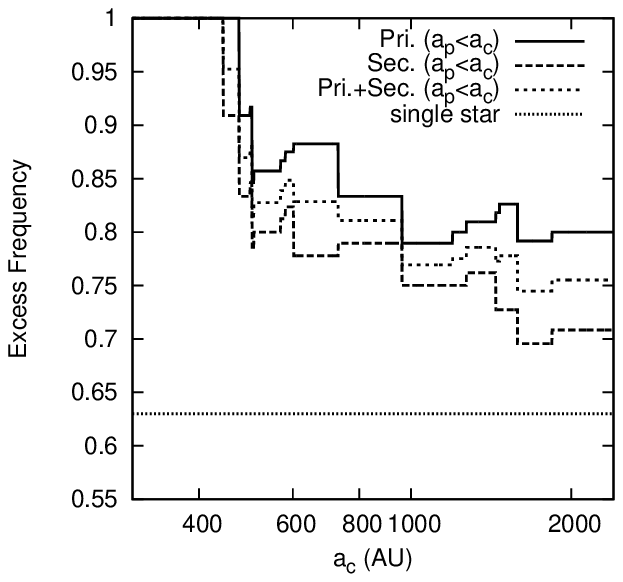}
\end{center}
\caption{
This figure shows the significance of EF differences between {\it closer binary systems} ($a_p < a_c$) and {\it single stars} without the calibration on possible tight companions described in the text.
{\it Left}: Fisher's $p$-value {\it vs.} $a_{\rm c}$ plane for primary only, secondary only, and primary+secondary sources ({\it solid}, {\it dashed}, and {\it dotted} lines, respectively).
The EF difference between the binary sample with $a_{\rm p}<a_{\rm c}$ and single stars was assessed by Fisher's exact test.
The {\it horizontal-dotted} and {\it dashed-dotted} lines indicate the 2$\sigma$ and 3$\sigma$ levels, respectively.
As in Fig.~\ref{fig:fishercvsw}, the $p$-value is minimized at $\sim$450~AU, where the difference in EF between binary systems with $a_{\rm p}<a_{\rm c}$ and single stars becomes most significant ($>$2$\sigma$ for each component and $>$3$\sigma$ for primary+secondary sources).
{\it Right}: The EF of binary systems and single stars.
The {\it horizontal-dotted} line indicates the ${\rm EF}_{\rm single}$ (63\%).
\label{fig:fishercvsswithout}}
\end{figure}

The difference in EF between single and binary systems was investigated first using this whole sample of single stars. Out of 99 sources, 62 show excess ($[3.6]-[8.0]>0.6$); i.e., ${\rm EF}_{\rm single}$ is 63\% (62/99).
This ${\rm EF}_{\rm single}$ was then compared with the EF of binaries with separations of $a_{\rm p}<a_{\rm c}$, again using Fisher's exact test. 
Figure~\ref{fig:fishercvsswithout} shows the results for the three subpopulations (primary, secondary, and both). 
The {\it left} and {\it right} panels present the Fisher's $p$-value and EF, respectively, as functions of $a_{\rm c}$. 
The most significant difference appears at $a_{\rm c}\simeq450$~AU, because the EF of primary and secondary sources separated within $\sim$450~AU is unity as described in Section~\ref{sec:efvssep}. 
This EF of 
each component is significantly ($>$2$\sigma$) higher than that of single stars of 63\%. For the entire binary sample including $both$ primary and secondary sources, Fisher's exact test gave the significance value of $>$3$\sigma$.
This suggests that, if the companions are separated by $\sim$280--450~AU, dust emission occurs more frequently in binaries than in single stars. 

\begin{figure}
\begin{center}
\FigureFile(78mm,80mm){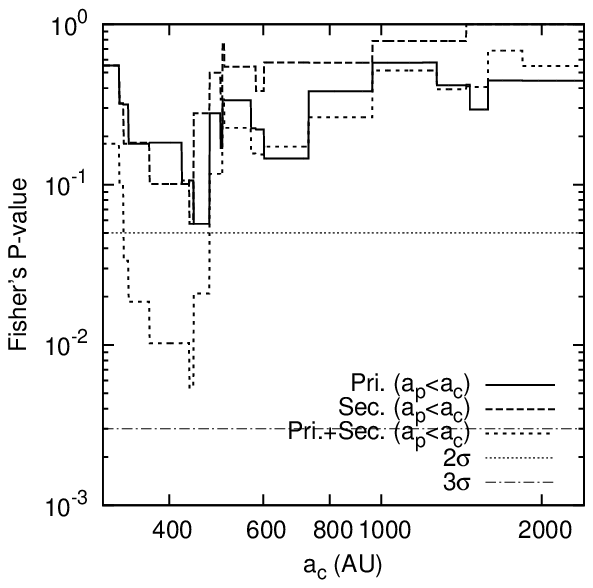}
\FigureFile(80mm,80mm){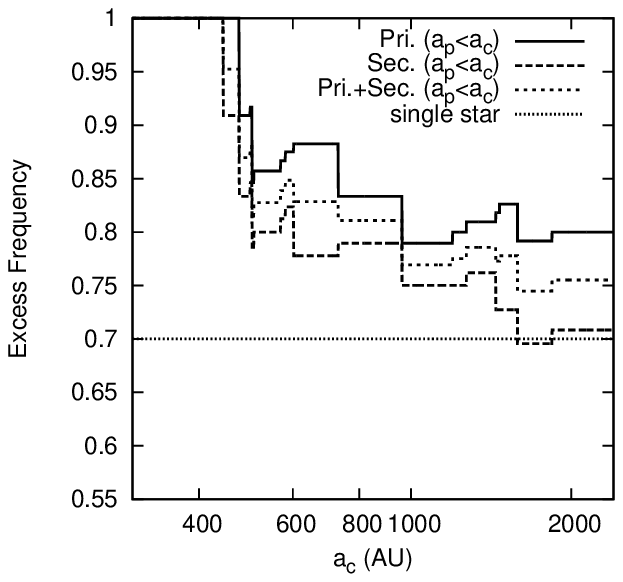}
\end{center}
\caption{
The figure is the same as Figure~6, but the comparison was performed between {\it closer binary systems} ($a_p < a_c$) and {\it single stars with calibration on putative tight companions}. 
{\it Left}: Fisher's $p$-value {\it vs.} $a_{\rm c}$ plane.
The {\it horizontal-dotted} and {\it dashed-dotted} lines indicate the 2$\sigma$ and 3$\sigma$ levels, respectively.
The $p$-value is minimized at $\sim$450~AU, where the difference in EF between binary systems with $a_{\rm p}<a_{\rm c}$ and single stars becomes most significant (2--3$\sigma$ for primary+secondary sources).
{\it Right}: The EF of binary systems and single stars.
The {\it horizontal-dotted} line indicates the ${\rm EF}_{\rm single}$ (70\%).
\label{fig:cali_FE}}
\end{figure}

However, since the 61 single stars in Ophiuchus have been less studied in high-angular-resolution compared to the stars in Taurus, the sample may include undetected close binaries.  
Considering a possible underestimate of ${\rm EF}_{\rm single}$, we attempted to calibrate ${\rm EF}_{\rm single}$  assuming the same fraction of close binaries ($\eta_{\rm cb}$) and the rate of disk-bearing ones ($\eta_{\rm cb;disk}$) in Ophiuchus as in Taurus. 
\citet{kra12} found that about 30\% of their Taurus targets turned out to be binaries with 1--40~AU separations, and one third of the binaries showed the signatures of disks ($\eta_{\rm cb}=0.3$, $\eta_{\rm cb;disk}=1/3$) . 
Note that the presence of a disk was inferred through the excess emission in 2--8~$\mu$m in their study. 
If the EF of the single stars in Ophiuchus (hereafter $\rm{EF}_{\rm{s;oph}}$) was determined with the same $\eta_{\rm cb}$ and $\eta_{\rm cb;disk}$, it yielded $\rm{EF}_{\rm{s;oph}} = 0.659$. 
Thus, the number of single stars was estimated as 42 for Ophiuchus, in which 28 were considered to have excess emission. 
Combining the Taurus sample with the calibrated one for Ophiuchus, ${\rm EF}_{\rm single}$ was found to be 70\%. 
We conducted Fisher's exact test using this sample. 
The Fisher's test demonstrated that the EF for each primary and secondary of the binaries is still significantly higher than that of single stars of 70\% if the binary separation is $a_{p}\sim$280--450~AU  ($\sim$2$\sigma$; Figure~\ref{fig:cali_FE}). The significance of the difference is 2--3$\sigma$ when estimated for the binary sample including both primary and secondary sources (see Figure~\ref{fig:cali_FE}). 
Therefore, we conclude the longer lifetime for binaries in this separation range. 

\section{Discussion}\label{sec:discussion}
\subsection{EF distribution as a function of disk outer radius}
The binary systems with projected separations, $a_p$, closer than $\sim$450~AU ($EF \simeq 100$\%) exceeded both wider systems and single stars at the 2--3$\sigma$ significance level in EF.
In binary systems, the outer edge of the circumstellar disk is predicted to be truncated and prevented from expanding freely by a companion star (e.g., \cite{art94}).  
This truncated radius, $R_{\rm out}$, strongly depends on the semi-major axis of the  binary system, $a$.
On the other hand, disk dispersal timescale is probably governed by multiple stellar and disk properties, but at least it should be linked to the size of the disk. Therefore, the relationship between EF and $a_{\rm p}$ can be interpreted by $R_{\rm out}$. 
We constructed a model of EF distribution  as a function of $R_{\rm out}$, EF$_{\rm model}(R_{\rm out})$, to explore whether the characteristic disk size exists to explain the observed high EF at the projected separations of 280--450~AU.
In the model, we divided the radial regions into three sections and assume a constant EF within one section. 
The radial ranges of the three regions and the EFs are treated as variables. 
To figure out the appropriate ranges of EF$_{\rm model}$ and $R_{\rm out}$ for this model, we calculated EF$_{\rm model}$($a_{\rm p}$) through EF$_{\rm model}(R_{\rm out})$ and compared it with the observed one, hereafter denoted as EF$_{\rm obs}(a_{\rm p})$. 

\begin{figure}
\begin{center}
\FigureFile(80mm,80mm){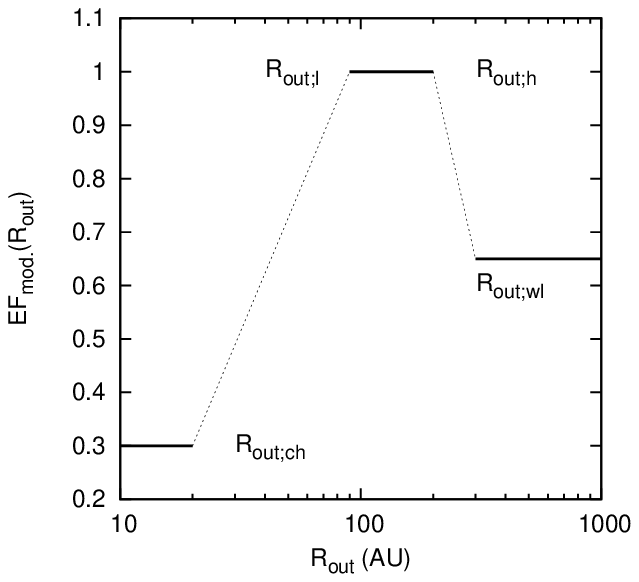}
\end{center}
\caption{
Schematic of the EF$_{\rm model}(R_{\rm out})$ model. 
This figure shows one example case of three EFs.
$R_{\rm out;ch}$ expresses the outer radius of the innermost EF region, where the EF is set to be  similar as observed in binaries separated by less than $\sim$40 AU (\cite{cie09}). 
$R_{\rm out;l}$ and $R_{\rm out;h}$ represent the inner and outer limit of the high EF region, respectively.
Introducing these two variables yields a ``peak'' in the EF distribution.
$R_{\rm out;wl}$ represents the inner boundary of the outermost EF region. 
The three regions are linearly connected ({\it dotted} lines in the figure).
\label{fig:efmodvsrout}}
\end{figure}

The relationship between $a$ and $a_{\rm p}$ is given by $a/a_{\rm p}=(1-e^2)[1-\cos^2(\theta-\phi)\sin^2i]^{0.5}/(1+e\cos\theta)$, where $\theta$, $\phi$, $i$, and $e$ denote the anomaly, reference angle, inclination, and eccentricity, respectively.
To derive the $a/a_{\rm p}$ distribution, the eccentricity distribution is required.
According to recent studies, the eccentricity distribution is relatively uniform (e.g., \cite{rag10,dup11}), and high-eccentricity ($e\gtrsim0.8$) binaries are absent. 
Therefore, we assume a uniform eccentricity distribution ranging from $e=0$ to 0.7.
Using this eccentricity distribution, a peak appears at $a/a_{\rm p}=1$ in the $a/a_{\rm p}$ distribution. 
In the previous section, we measured EF$_{\rm obs}$$(a_{\rm p}\simeq280$--$450~{\rm AU})=100$\% and EF$_{\rm obs}$$(a_{\rm p}>450~{\rm AU})\simeq70$\%; therefore, we expect that EF$_{\rm model}$$(a\simeq280$--$450~{\rm AU})=100$\% and EF$_{\rm model}$($a>450~{\rm AU})\simeq70$\% would give a EF distribution consistent with the observations. 
In addition, ${\rm EF}_{\rm model}(a\lesssim40~{\rm AU})\sim20$--40\% is assumed based on the report by \citet{cie09}. 
Adopting a truncation radius of $R_{\rm out}=0.337a$ found in the systems of mass ratio $q\equiv M_{\rm s}/M_{\rm p}=1$ with  low-eccentricity  \citep{pic05}, the EF can be estimated as a function of $R_{\rm out}$ over three separation regions; $R_{\rm out}\lesssim10$ AU, $R_{\rm out}\sim100$--150 AU, and $R_{\rm out}\gtrsim150$~AU. 
From the above arguments, 
the ${\rm EF}_{\rm model}(R_{\rm out})$'s in the three separation regions are varied around $\sim$20, 100, and 70\%, respectively.
Between the three regions, the EF is set by linear interpolation, as shown in Figure~\ref{fig:efmodvsrout}.
The inner  and outer limits of the radial region of ${\rm EF}_{\rm model}(R_{\rm out})\simeq1$ are denoted as $R_{\rm out;l}$ and $R_{\rm out;h}$, respectively. 
We also define $R_{\rm out;ch}$ and $R_{\rm out;wl}$ as the outer boundary of the ${\rm EF}_{\rm model}(R_{\rm out})\simeq0.2$ and the inner one of the ${\rm EF}_{\rm model}(R_{\rm out})\simeq0.7$ region, respectively.
From the relationship between EF$_{\rm model}$ and $R_{\rm out}$ determined from free parameters \{$R_{\rm out;l}$, $R_{\rm out;h}$, $R_{\rm out;ch}$, $R_{\rm out;wl}$\} and the EFs in the three separation regions, we can derive \{$a_{\rm p}$, $R_{\rm out}$, EF$_{\rm model}(a_{\rm p})$\} from randomly selected binary orbital parameters $\{a, e, i, \theta, \phi\}$.
The calculation assumes a uniform distribution of semi-major axes.
This Monte Carlo approach allows us to compare the ${\rm EF}_{\rm model}(a_{\rm p})$ and ${\rm EF}_{\rm obs}(a_{\rm p})$ in order to determine the parameters related to the ${\rm EF}_{\rm model}(R_{\rm out})$.

\begin{figure}
\begin{center}
\FigureFile(80mm,80mm){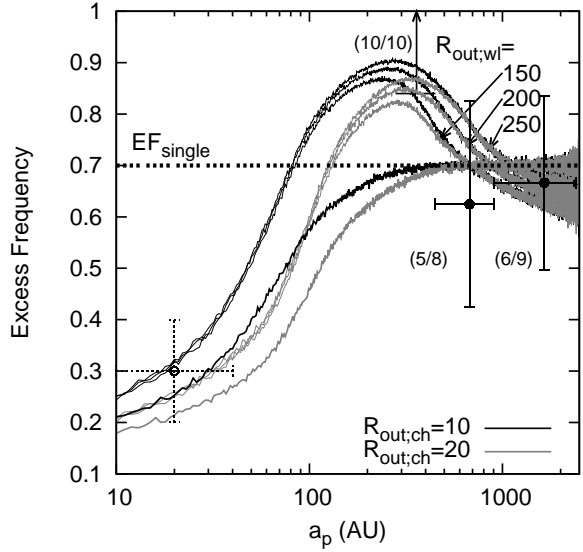}
\end{center}
\caption{
EF$_{\rm model}(a_{\rm p})$ derived from EF$_{\rm model}(R_{\rm out})$. 
{\it Black} and {\it gray solid} lines denote $R_{\rm out;ch}=10$ and 20~AU, respectively. 
In the peak model, the EF at wide binary separations ($a_{\rm p}\gtrsim100$) depends on $R_{\rm out;wl}$.  
The selected parameter values are indicated in this figure.
If no peak is present in the distribution, the EF is almost flat at wide separations. 
The observational EF data for primary sources are shown for comparison. 
Error bars represent the 1$\sigma$ confidence interval estimated from the binomial distribution.
As reference, the EF of a closely spaced binary ($<$40 AU, from \cite{cie09}) is plotted with dotted error bars. 
The {\it dotted horizontal} line represents the EF of a single star (see~Section~\ref{sec:single}).
\label{fig:efmodvsap}}
\end{figure}

Figure~\ref{fig:efmodvsap} shows the results of the Monte Carlo computations obtained with $R_{\rm out;l}$, $R_{\rm out;h}$, $R_{\rm out;ch}$, and $R_{\rm out;wl}$ set to 30--40, 100, 10--20~AU, and 150--250 AU, respectively.
The EFs of the three characteristic regions are 10\% ($<$$R_{\rm out;ch}$), 100\% ($R_{\rm out;l}$--$R_{\rm out;h}$), and 70\% ($>$$R_{\rm out;wl}$).  
Although the flat EF regions shown in Figure~\ref{fig:efmodvsrout} are smoothed by the projection effect, the EF remains high around $a_{\rm p}\simeq280$--450~AU. 
In other words, the observed high EF region was reproduced when introducing this ``peak'' of EF into the model distribution. 
The model results $without$ peaks are shown in the figure for comparison. 
For the model without a peak, we set EF$_{\rm model}$($>$$R_{\rm out;l}$) = 70\% so that the EF$_{\rm model}$($a_p$) at larger $a_p$ approaches the observed EF for wider binaries and single stars. 
We also set EF$_{\rm model}$($R_{\rm out;ch}$)=10--20\% so that the EF of the inner 40 AU is consistent with the observed one, 20--40\% \citep{cie09}. 
These results disagree with the observational data around $a_{\rm p}\simeq280$--450~AU (Figure 9).

\begin{figure}
\begin{center}
\FigureFile(80mm,80mm){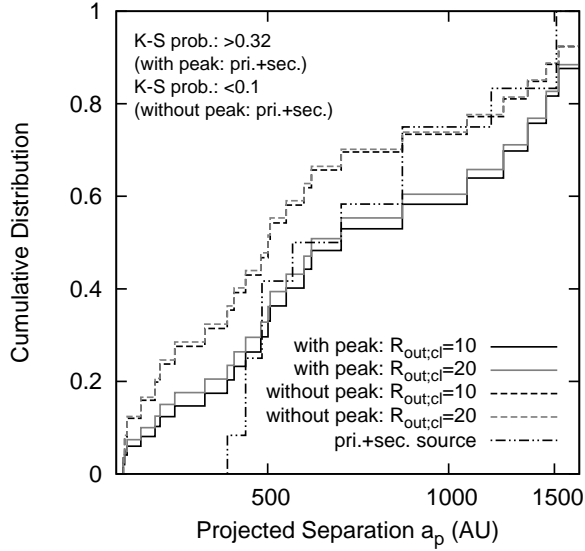}
\end{center}
\caption{
Cumulative distributions of the projected separation, $a_{\rm p}$.
As demonstrated by the K-S tests, if the ``peak'' is excluded from the ${\rm EF_{\rm model}}(R_{\rm out})$ distribution, 
discrepancy appears between the predicted and the observed separation-dependent EF distribution.
{\it Solid} and {\it dashed} lines show the results with and without the peak in the model, respectively.
{\it Black} and {\it gray} colors denote the calculations assuming $R_{\rm out;ch}=10$ and 20, respectively.
The observed distribution 
is indicated by 
the {\it dash-two-dotted} line. 
\label{fig:cdf}}
\end{figure}

The separation-dependent distribution was investigated individually for excess and non-excess sources, 
by a Kolmogorov-Smirnov (K-S) test. 
We iteratively sampled $\sim$$10^{4}$ sources from the ${\rm EF}_{\rm model}(R_{\rm out})$ at the same $a_{\rm p}$ as the observed binary, and counted the number of excess or non-excess objects at each $a_{\rm p}$. 
In this way, our model predicts how many sources show and do not show excess at $a_{\rm p}$ and their cumulative distribution functions (CDF).
According to the K-S test, the model predicts the separation-dependent distribution of both primary and secondary excess sources, regardless of whether the peak is included in the model.
However, for sources that do not show excess, the model predictions disagree with observations when the peak is excluded. 
As shown in Figure~\ref{fig:cdf}, $p\gtrsim$0.32 is obtained by the K-S test where $p$ is the probability that the observed binary sources (primary plus secondary) are drawn from the same parent sample as the model {\it with} an EF peak.  On the other hand, $p$-value is $\lesssim$0.1 for the model without a peak, suggesting disagreement with the observations. 
Therefore, although it is not statistically significant, the peak-ed ${\rm EF}_{\rm model}$ is more consistent with the observed result than the flat ${\rm EF}_{\rm model}$. 
We discuss the interpretation of the EF peak seen in $a_{\rm p}\simeq280$--450~AU  and attempt to explain why the EF of this separation range is higher compared to that of the single stars in the next section.

\subsection{Mechanism of dust dispersal in binary systems}
The EFs of primary and secondary sources in our sample separated by $a_{\rm p}\simeq280$--450~AU are $100^{+0}_{-17}$\% and $91^{+8}_{-18}$\%, respectively.
At wider separations, the EF reduces to that of single stars ($70\pm5$\%).
On the other hand, the EF of close binaries ($a_{\rm p}\sim10$--40~AU) has been previously determined as $\sim$20--40\% \citep{cie09}.
In the previous section, we expressed the separation-dependent EF in terms of $R_{\rm out}$. 
The EF significantly decreases, relative to that of single stars, when $R_{\rm out}\lesssim10$~AU, increases when $R_{\rm out}\sim30$--100~AU, and matches the single star value at $R_{\rm out}\gtrsim150$~AU.
This has the implication: {\it the time scale of dust dissipation might be prolonged in disks of characteristic radius ($\sim$30--100~AU)}.
This disk size is smaller than the size estimated for single stars, reported as $\sim$200~AU (e.g., \cite{and07}) and $\sim$100--1100~AU (\cite{sch09} and references therein).
Given that the EF is reduced in wide binary systems ($\gtrsim$450~AU, thus $R_{\rm out}\gtrsim150$~AU), we speculate that extended disks (greater than $\sim$30--100~AU) undergo significant mass loss in the outer disk region.

Various mechanisms of gas and dust removal have been proposed, including viscous evolution, planetesimal growth, photoevaporation, disk wind induced by magnetorotational instability, and sculpting by planets (e.g., \cite{wil11,suz10}).
In viscous evolution, the timescale of dissipation is given by $t\propto R^2_{\rm out}$.
The $R_{\rm out}$ of binary systems depends upon the semi-major axis $a$, being approximated by $\sim$$a/3$ (e.g., \cite{art94,pic05}).
Therefore, the peak in the EF distribution is difficult to reproduce assuming viscous evolution alone, since more time is required to disperse the dust disk as the binary separation (hence the disk size) increases.

Disk photoevaporation is a well-known mechanism and may successfully explain the actual disk dispersal  (e.g., \cite{hol94,cla01,ale07}). 
Ionizing flux from the central star creates a disk thermal wind and material beyond the critical radius can escape the disk as the gravitational potential weakens. On the other hand, viscous accretion continues, leaving an inner hole within the critical radius due to the lack of sufficient mass supply from the outer region and the short timescale of viscous process at small radii. 
\citet{arm03} considered the photoevaporation in their study of disk evolution in binary systems. In their modeling work, the disk lifetime in a binary depended on its separation, and the maximum lifetime  was predicted at $\sim$15--100~AU, depending on the model which they used. They speculated that binary systems with those intermediate separations have less area where the disk mass can be lost by photoevaporative wind and moderately counteract the disk dispersal. 
This earlier prediction cannot be verified by the results presented here because we cannot ascertain whether the sources with no IRAC excess really lack outer disks without analyzing the longer wavelengths data. However, it is worth pointing out that the EF distribution has a peak in the intermediate separation range also in our study. 
This similarity between the observed result and the theoretical prediction implies that photoevaporation has a possibility of being one of the effective mechanisms controlling the disk evolution in binary systems.

Disk lifetime is related also to disk mass. 
If we assume the disk dispersion  described above, among the binaries of $a_p\gtrsim280$~AU, the closer binaries ($a_p$$<$ 450~AU) should have heavier disks than the wider binaries ($a_p$$>$450 ~AU). 
However, such a trend has not been found so far. Binaries separated by $>$280~AU are speculated to have the same degree of millimeter luminosity, in other words, the same disk mass as single stars 
\citep{har12}. Unfortunately, it is still difficult to statistically confirm the change in disk mass within a narrow separation range of $\sim280$--500~AU. 

There is another possibility to interpret the dependence of EF on binary separation. Systems with wider separations tend to be triples or quadruples. If the 450~AU represent the boundary beyond which binaries become sufficiently wide that they start to have additional higher-order components, and they harbor close companions, the decreasing of EF can be accounted for by an increasing of existence of multiple systems. In order to clarity this possibility, a larger sample is needed for wide binaries for which the presence of close companions were thoroughly investigated.

\section{Summary}
By using archival data obtained by {\it Spitzer}/IRAC, infrared excess emission was investigated for 27 binary systems in two nearby star-forming regions, Taurus and Ophiuchus.
Our sample consisted of binaries with projected separations of $2\farcs0$--$17\farcs0$ ($\simeq$280--2400~AU), enabling us to study circumstellar disks associated with individual binary components.

After excluding Class I and objects that cannot be young, spatially resolved photometry yielded the following results for 24 binaries or 51 primary and secondary sources. 
\begin{enumerate}
\item 
The excess frequency (EF) of $all$ primary and secondary stars based on $[3.6]-[8.0]$ color is indistinguishable from that of single stars ($70\pm5$\%).
This result is consistent with past studies in which the disk frequency of wide binary systems ($\gtrsim$100~AU) is not significantly different from that of single stars.
The sources showing excess emission at 8.0~$\mu$m also exhibits excess at the shorter wavelengths. 
Conversely, sources without excess at 8.0~$\mu$m are photospheric at the other IRAC wavelengths.
The exceptions are three sources that show excess at 8.0~$\mu$m but not in one or more shorter wavebands.
\item In four systems out of 24 ($17^{+11}_{-8}$\%), excess emission is shown by either the primary or secondary component (designated ``mixed'' systems).
In one of these mixed systems, excess is found only for the secondary star.
No significant preference for either the circumprimary or circumsecondary disk to disperse is  found.
However, the rareness of mixed systems suggests that dust dispersal is strongly correlated between the two disks, even at the binary separation exceeding $\gtrsim$280~AU.
\item 
The EFs of closely and widely spaced binary systems are different, consisting of the two groups  divided by $a_{\rm p}\sim$450~AU. 
For the entire sample (including both primary and secondary sources), this difference is  significant at the 2--3$\sigma$ level.
The EF of primary and secondary sources separated by less than 450~AU was $100^{+0}_{-17}$\% and $91^{+8}_{-18}$\%, respectively.
This high EF was compared with that of single stars within a similar range of  luminosity  and spectral type (i.e., similar age and mass). 
The Fisher's exact test shows that the maximum EF ($\sim$100\%) is higher than that of single stars with a significance of $\gtrsim$2--3$\sigma$ for three kinds of binary sample consisting of primary or secondary or both components. 
\end{enumerate}

Comparing the predicted separation dependence of the EF distribution with the observational data, the model requires a high EF region (``peak'') around $R_{\rm out}\sim$30--100~AU, as confirmed by the K-S test.
This disk size is smaller than the typical one estimated for single stars, implying that the dissipation of moderately truncated circumstellar disks is prolonged.
The EF of widely-spaced binary systems, presumably with larger circumstellar disks, is indistinguishable from that of single stars. 
Our result is consistent with the prediction that extended disks may suffer from significant mass loss in the outer regions by photoevaporation \citep{arm03}, suggesting that photoevaporation contributes to the evolution of circumstellar disks in binary systems. 

\bigskip

We are grateful to the referee for valuable comments to improve the paper. This work is based on archival data obtained by the Spitzer Space Telescope, which is operated by the Jet Propulsion Laboratory, California Institute of Technology, under a contract with the National Aeronautics and Space Administration (NASA). Support for this work was provided by NASA. 
This publication utilizes data products from the Two Micron All Sky Survey, which is a joint project of the University of Massachusetts and the Infrared Processing and Analysis Center/California Institute of Technology and is funded by the NASA and the National Science Foundation.

\appendix
\section{Notes on individual binaries}\label{sec:app}
\subsection{2MASS~J16263682-2415518~B}
This binary system is located in Ophiuchus. The secondary star begins to exhibit significant excess ($0.48\pm0.14$~mag) at 8.0~$\mu$m. 
The binary separation is very wide at $9\farcs08$ \citep{duc07}, and the spectral type of the secondary star is estimated as M5 \citep{wil05}.

The SED fitting was performed to the flux density in $R$, $I$, and $J$ bands obtained from USNO-B1 and 2MASS Catalog.
Significant excess emission was confirmed at 8.0~$\mu$m, but was not identified at 5.8~$\mu$m. 
However, the possible excess at 5.8~$\mu$m might have been underestimated because the photometric values at 3.6 and 4.5~$\mu$m are smaller than the fitted photosphere. The measured color of $[4.5]-[5.8]$ ($0.49\pm0.2$~mag) is in fact significantly higher than the photospheric color for an M5-type star (0.04 mag; \cite{luh10}), thus excess emission likely begins at 5.8~$\mu$m. 
The discrepancy between the photometry and the fitted photospheric values at 3.6 and 4.5~$\mu$m could arise from the poor SED fitting due to the faintness of the object in the $R$ band (19.2~mag).
Therefore, according to its color, we consider that this object has an onset of the excess at $\sim$5.8--8.0~$\mu$m.

The source was detected at 24~$\mu$m in the c2d survey, and the photometric value of $5.71\pm0.19$~mag was provided in the catalog. 
The $[8.0]-[24]$ color is then $3.46\pm0.24$~mag, which is significantly higher than the photospheric one for an M5-type star (0.23~mag; \cite{luh10}).
Thus, we again confirmed the existence of circumsecondary dust, and the occurrence of dust dispersion in the inner region of the disk.

\subsection{V710~Tau~A}
The binary separation of V710~Tau is $3\farcs0$ \citep{whi01}. 
The spectral types of the primary and secondary stars are estimated as M0.5 and M2, respectively \citep{whi01}. 
In V710 Tau, excess emission is found only for the primary source.
The subtraction of the photospheric model derived from the SED fitting resulted in significant excess at 5.8 and 8.0~$\mu$m, with values of $0.37\pm0.14$ and $0.94\pm0.14$~mag, respectively.
On the other hand, no excess was observed at 3.6 and 4.5~$\mu$m. 
The $[4.5]-[5.8]$ color of $0.67\pm0.2$~mag is significantly higher than the photospheric color of 0.2 mag, also suggesting that excess emission begins from 5.8~$\mu$m.
In other wavebands, the primary star shows a small excess in $L$ band ($K-L=0.44$; \cite{whi01}) and a substantial one in $N$  ($K-N=3.32$; \cite{mcc06}).  
The significant excess starting from 5.8~$\mu$m is not inconsistent with these previous studies.

The flux density at 24~$\mu$m was obtained as 236~mJy for the whole system (primary plus secondary) \citep{wah10}.
The magnitudes measured at 8.0 and 24.0~$\mu$m for the system are $6.30\pm0.20$~mag and $3.71\pm0.11$ mag, respectively, then the color of $[8.0]-[24]$ is $2.59\pm0.23$~mag. This is significantly larger than an averaged-photospheric color of an M-type star of 0.22~mag (\cite{luh10}).
Because the emission at 24~$\mu$m cannot be spatially resolved, it is  difficult to conclude whether the origin of the emission is circumprimary, circumsecondary, 
both of them, or circumbinary disks.
However, considering the separation of $\sim280$~AU, the emission at 24~$\mu$m is possibly not from the inner edge of a circumbinary disk.

\subsection{JH~223~A}
The binary JH~223 belongs to the Taurus region. 
The primary star of this system  shows the significant excess at 4.5~$\mu$m.
The binary separation is small ($2\farcs07$; \cite{kra11}), and the spectral types of its primary and secondary are estimated as M2 and M6.5, respectively \citep{kra07}.

The excess emission at 3.6, 4.5, 5.8, and 8.0~$\mu$m were $-0.04\pm0.03$, $0.33\pm0.04$, $0.61\pm0.04$, and $1.03\pm0.04$~mag, respectively, estimated by subtracting the photospheric magnitudes obtained from the SED fitting. This indicates that the onset of the excess is at around 4.5~$\mu$m.
The resultant $\chi^{2}/{\rm d.o.f.}$ values for the PSF fitting photometry were 1.0, 0.4, 0.7 and 1.0 at the four wavelengths, respectively, implying that the fitting was reasonably performed. 
The H$\alpha$ emission have  identified  this source as a Weak-lined T Tauri Star (\cite{neu95} and references therein), suggesting that this source lacks an optically-thick, inner disk with active accretion, but may possess an optically-thick, exterior disk.

The photometry at 24~$\mu$m was previously obtained as $5.13\pm0.04$~mag for the primary and  secondary sources \citep{reb10}.
The measured magnitudes at 8.0 $\mu$m of the system is $7.72\pm0.20$~mag and the $[8.0]-[24]$ color is calculated to be $2.59\pm0.20$~mag, which is significantly higher than an averaged-photospheric value of an M-type star of 0.22~mag \citep{luh10}.
As in the case of V710 Tau, although it is hard to conclude the origin of the excess emission, the emission at 24~$\mu$m may not be caused by the circumbinary disk. 

\end{document}